\documentclass[12pt]{article}
\usepackage{graphicx}
\usepackage{epsfig}
\usepackage{latexsym}
\usepackage{latexsym}
\usepackage{amssymb}
\newcommand{\labell}[1]{\label{#1}}
\newcommand{\reef}[1]{(\ref{#1})}

\DeclareSymbolFont{AMSb}{U}{msb}{m}{n}
\DeclareMathSymbol{\IN}{\mathbin}{AMSb}{"4E}
\DeclareMathSymbol{\IZ}{\mathbin}{AMSb}{"5A}
\DeclareMathSymbol{\IR}{\mathbin}{AMSb}{"52}
\DeclareMathSymbol{\Q}{\mathbin}{AMSb}{"51}
\DeclareMathSymbol{\II}{\mathbin}{AMSb}{"49}
\DeclareMathSymbol{\IC}{\mathbin}{AMSb}{"43}
\DeclareMathSymbol{\IP}{\mathbin}{AMSb}{"50}
\DeclareMathSymbol{\IH}{\mathbin}{AMSb}{"48}
\DeclareMathSymbol\IA{\mathalpha}{AMSb}{"41}
\DeclareMathSymbol\IS{\mathalpha}{AMSb}{"53}

\def\Q{{\cal Q}}

\begin{document}

\hbox{$\phantom{.}$}

\bigskip
\bigskip
\begin{center}
   {\Large \bf Rotating Black Holes, Closed Time--Like Curves,}

\bigskip

 {\Large \bf Thermodynamics, and  the  Enhan\c con Mechanism}




\end{center}

\bigskip
\bigskip
\bigskip
\bigskip

\centerline{\bf Laur J\"arv$^{a}$ and 
Clifford V. Johnson$^b$}

\bigskip
\bigskip

\centerline{\it ${}^{a}$Theoretisch--Physikalisches Institut}
\centerline{\it Friedrich--Schiller--Universit\"at Jena}
\centerline{\it Max--Wien--Platz 1}
\centerline{\it  D-07743 Jena, Germany}

\bigskip
\bigskip

\centerline{\it ${}^{a,b}$Centre for Particle Theory} 
\centerline{\it Department of Mathematical Sciences}
\centerline{\it University of Durham}
\centerline{\it Durham, DH1 3LE, U.K.}

\bigskip
\bigskip

\centerline{$\phantom{and}$}

\bigskip

\centerline{\small \tt
l.jaerv@tpi.uni-jena.de,
c.v.johnson@durham.ac.uk}

\bigskip
\bigskip
\bigskip

\begin{abstract}
  \bigskip We reconsider supersymmetric five dimensional rotating
  charged black holes, and their description in terms of 
  D--branes.  By wrapping some of the branes on K3, we are able to
  explore the role of the enhan\c con mechanism in this system. We
  verify that enhan\c con loci protect the black hole from violations
  of the Second Law of Thermodynamics which would have been achieved
  by the addition of certain D--brane charges. The same charges can
  potentially result in the formation of closed time--like curves by
  adding them to holes initially free of them, and so the enhan\c con
  mechanism forbids this as well.  Although this latter observation is
  encouraging, it is noted that this mechanism alone does not
  eliminate closed time--like curves from these systems, but is in
  accord with earlier suggestions that they may not be manufactured,
  in this context, by physical processes.
\end{abstract}

\newpage
\baselineskip=18pt
\setcounter{footnote}{0}


\section{Introduction  and Conclusions}

As we increase our ability to describe the more exotic types of
physics which branes can exhibit (and here we have in mind their
ability to change shape, dimension, and other key aspects of their
character) we confirm our suspicions that they are part of a fruitful
avenue of research into the basic nature of the correct description of
spacetime physics and whatever replaces it at the most fundamental
level.

A particular example that we have in mind was uncovered last
year\cite{jm} by reconsidering the case of extremal five dimensional
black holes and their microscopic description\cite{Strominger:1996sh}
in terms of D1--branes and D5--branes wrapped on K3$\times S^1$. The
intriguing piece of physics observed in that study was the fact that
while an analysis of the
entropy as a function of the R--R charges suggests that an 
approach to the horizon of an additional wrapped D5--brane could
decrease the entropy and hence violate the Second Law of
Thermodynamics, a further analysis in the light of the results of
ref.\cite{jpp} shows that the enhan\c con mechanism ---which forces
the brane to delocalise, spread out, and cease its approach to the
black hole at a specific radius--- prevents this from happening. This
mechanism extends to the entire family of D5--D1 bound states with
particular charge assignments such that they have the potential to
allow Second Law violations, while allowing other types through
---there is an enhan\c con locus for each type of bound state; above
the horizon for violators and below it for law abiders.

This result is both satisfying and intriguing, since in ordinary
thermodynamics, the preservation of the Second Law is understood as a
course--grained statistical outcome, while in this case, since
D--branes (in terms of which the black holes find their microscopic
description\cite{Strominger:1996sh}) are the smallest possible objects
carrying the R--R charges\cite{gojoe}, the Second Law is kept
inviolate by a macroscopic supergravity filter which can discriminate
at the level of the microscopic constituents. This seems to be a new
sharp phenomenon connecting the microscopic to the macroscopic,
deserving further investigation in order to be better understood.

The principal reason why we get to address this system quite cleanly
in the above terms is because it is a BPS system. Therefore, we can
separate it into small non--interacting BPS pieces and bring each
piece up to it slowly and hence perform adiabatic changes to the
thermodynamic quantity of interest, the entropy\footnote{It is
  straightforward to see that this sort of reasoning also works very
  well for the four dimensional Reissner--Nordstr\"om black holes
  which admit a simple description involving D1--branes and D5--branes
  (and possibly NS5--branes) wrapping K3\cite{fourdee}, as was
  confirmed in ref.\cite{neil}. The same sort of BPS processes can be
  found in those systems too.}, without the encumbrance of having to
worry about the thermodynamic properties of radiative processes, {\it
  etc}.  The purpose of this short note is to extend the result of
ref.\cite{jm} to a class of rotating charged black holes, taking
advantage of the fact that although naively the more complicated case
of a rotating black hole would seem to admit no BPS embedding into
string theory, there is indeed such a solution, found in
ref.\cite{Breckenridge:1996is}, which owes its BPS nature\footnote{All
  supersymmetric solutions of minimal supergravity in five dimensions
  were found in ref.\cite{Gauntlett:2002nw}.} to an excellent
conspiracy of features in five dimensions, discussed in
ref.\cite{Gauntlett:1998fz}, and further in
refs.\cite{Gibbons:1999uv,Herdeiro,townsend}. We show here that we can
carry out a very similar analysis for the rotating black hole to that
carried out for the static case, and observe that the potential
violations of the Second Law using the adiabatic addition of
constituent parts of the ``incorrect'' charges is again avoided by the
enhan\c con mechanism.

The bottom line, of course, is that the area of the black hole's
horizon, to which the entropy is proportional\cite{Bekenstein:ur}
cannot decrease.  This follows from an appropriate version of the area
theorem\cite{hawkarea}, adapted to the case in question.  Such
theorems follow from a weak energy condition, and may also be thought
of as a corollary of the  Cosmic Censorship
Principle\cite{penrose}.  (See refs.\cite{MTW:book,Wald:rg} for
further discussion.) Our goal here is not, therefore, to find
violations {\it per se}, but to study the novel mechanism by which
this particular situation involving branes wrapped on K3 manages to
protect the theorem. So cases which are in the same class of physics
as the $T^4$--wrapped situation ---branes with the ``correct''
charges--- will be irrelevant to the enhan\c con mechanism, and will
be covered by the area theorem in the usual way.  We have nothing to
add to the existing discussion for those cases.

As it is a while since many have thought about these models, we review
much of the essential material in the short sections~2, 3 and 4, which
also allow us to establish our notation and emphasise the crucial
differences between the K3 and $T^4$ cases.  In section~5, we exhibit
the basic enhan\c con locus and in section~6 we perform a D--brane
probe computation which enjoys a crucial cancellation due to the
particular form of its interaction with the background fields
representing the rotating solution. This cancellation is crucial for
forming the intuition about how to make a rotating BPS black hole,
{\it i.e.,} having no excess energy taking us away from BPS saturation.
In that section, we also show that there is a whole family of enhan\c
con loci, one type for each possible arrangement of R--R charges that
a D--brane probe can carry.  Section~7 exhibits a supergravity
excision computation which demonstrates that the enhan\c con shells
suggested by the probe computation really exist as solutions, and have
the required properties. Section~8 then considers the case of a black
hole with non--zero entropy, and shows that the class of probes which
have R--R charges which could potentially reduce the entropy if they
merge with the hole are stopped from reaching the horizon by the
enhan\c con mechanism. The remaining sorts of probes are harmless, and
the enhan\c con mechanism has nothing to say about their motion
outside the horizon.

We also note the following. While the ten dimensional geometry of the
brane configuration giving rise to the black hole in the five
dimensional supergravity upon reduction is entirely causal, the
reduction process leads to naked closed time--like curves (CTCs) in
the geometry if the angular momentum of the solution exceeds a certain
bound\cite{Breckenridge:1996is}\footnote{This is in contrast to the
  case of the generalisation\cite{Klemm:2000vn} of these solutions to
  the case of the analogous gauged five dimensional supergravity. In
  that case, their lift to ten dimensions still has 
  CTCs\cite{Caldarelli:2001iq}.}.  (See also
refs.\cite{Gauntlett:1998fz,Gibbons:1999uv,Herdeiro} for more
discussion. It was noted in ref.\cite{Gauntlett:2002nw} that CTCs seem
to be generic for a wide class of solutions in five dimensions.)  In
fact, such solutions are not ruled out by supersymmetry, and it is as
yet an unresolved question as to whether string theory has any
concrete mechanism for rendering CTCs more physically acceptable than
they appear to be in field theory.

We do not expect that the enhan\c con mechanism can be the tool by
which we find complete understanding of the role of CTCs since while
the former arises in this context from wrapping on K3, the latter have
nothing to do with the reduction on K3, and arise for the case of
$T^4$ as well. However, since the enhan\c con mechanism seems to be
intimately familiar with the physics of the entropy of the black
horizon formed upon reduction ---enough to be ``mindful'' of the
Second Law--- it is not unreasonable to wonder what it has to say
about the formation of the CTCs. We find that while it does not rule
out the existence of CTCs ---the expected result--- the same mechanism
which prevents the violation of the Second Law by ``wrong charge''
probes also prevents one from starting with a hole with no CTCs and
adding such branes in such as way as to form CTCs. This is in accord
with the work presented in ref.\cite{Gibbons:1999uv}, suggesting that
these CTCs may not be formed by physical processes.

\section{Five Dimensions}
\label{sec:5d}
Consider the following five dimensional action\cite{Hawking:bu}:
\begin{equation}
S_{(5)}=\frac{1}{16\pi G_5}\int d^5x \left(
  \sqrt{-g}\left[R-F^2\right]-\frac{2}{3\sqrt{3}}A\wedge F\wedge
  F\right)\ ,
\labell{sugra}
\end{equation}
where $A$ is a gauge field and $F$ is its field strength, combining
into a Chern--Simons term. This is the bosonic content of an ${\cal
  N}=2$ supergravity theory, which can be embedded into string theory
in a number of ways.  The metric of the rotating solution is written in
Einstein frame as:
\begin{eqnarray}
ds_{(5)}^2 &=& - H^{-2} \left( dt + \frac{J}{2 r^2} ( \sin^2\theta d\phi_1 - \cos^2\theta d\phi_2) \right)^2
 + H\left( dr^2 + r^2 d\Omega_3^2 \right)\ ,\nonumber\\
H&=&\left(1+\frac{r_0^2}{r^2}\right)\ ,
\labell{fivemetric}
\end{eqnarray}
where $\phi_1,\phi_2$ and $\theta$, ($0\leq\phi_{1,2}\leq 2\pi$,
$0\leq\theta\leq\pi/2$) are angles parameterising a round $S^3$, with
metric:
\begin{equation}
 d\Omega_3^2= d\theta^2 + \sin^2\theta ~d\phi_1^2 + \cos^2\theta
~d\phi_2^2\ ,
\end{equation}
Together with the  following gauge field:
\begin{eqnarray}
A&=&\left(H^{-1}-1\right)dt+\frac{J}{2r^2}H^{-1}\left(\sin^2\theta
  d\phi_1 - \cos^2\theta d\phi_2\right)\ ,
\end{eqnarray} the metric~\reef{fivemetric}
represents a very special solution with a regular event horizon
located at $r=0$. This solution is special since it is in fact a BPS
solution, brought about by the presence of the Chern--Simons term and
also a particular (anti) self--duality property of the gauge
field\cite{Gauntlett:1998fz,Gibbons:1999uv}.

One amusing feature of the solution is the fact that  although the
geometry has non--vanishing asymptotic angular momentum, the angular
velocity of the horizon is actually zero. As there is a negative
contribution to the angular momentum from the spacetime inside the
horizon, this vanishing is attributed to the cancellation of opposite
``dragging effects'' at the horizon\cite{Gauntlett:1998fz,townsend}.

The entropy of this solution can be easily computed by use of the
Bekenstein--Hawking relation\cite{Bekenstein:ur} to the
horizon area $\mathcal{A}$:
\begin{eqnarray}
S & = & \frac{\mathcal{A}}{4 G_{5}}  = 
 \frac{2 \pi^2}{4 G_{5}} \sqrt{r_0^6 -
 \frac{J^2}{4}}\ .
\labell{rotating-black-hole-entropy}
\end{eqnarray}

It is very interesting to note that this quantity vanishes for large
enough $J$. In fact, it can be seen that the geometry can develop
closed time--like curves if $J$ were to increase further.  For
example, picking either $\phi_1$ or $\phi_2$ (and calling it $\phi$),
an examination of the worst case behaviour of the metric for that
direction yields:
\begin{equation}
g_{\phi\phi}(r)=\frac{1}{(r^2+r_0^2)^2}\left(r^6+3r^4r_0^2+3r^2r_0^4+r_0^6-\frac{J^2}{4}\right)\ ,
\labell{timelike}
\end{equation}
showing that for $J^2>4r_0^6$, the closed loop parameterised by $\phi$
goes timelike above the horizon ({\it i.e.,} for $r>0$).

\section{Ten Dimensions}
The supergravity given in equation~\reef{sugra}, and the
solution~\reef{fivemetric} can be generalised, as there are more
independent gauge fields and a family of scalar fields which can be
switched on. These can then be seen to be fields arising from the
various geometrical choices to be made in embedding the supergravity
into ten dimensional string theory.

The five dimensional Einstein frame metric for the more general solution is\cite{Breckenridge:1996is}:
\begin{eqnarray}
&&\hskip-1cm ds_{(5)}^2 = - \left(H_1 H_5 H_P\right)^{-2/3}
\left( dt + \frac{J}{2 r^2} ( \sin^2\theta d\phi_1 - \cos^2\theta d\phi_2) \right)^2 
+ \left(H_1H_5H_P\right)^{1/3} \left( dr^2 + r^2 d\Omega_2^2 \right)\ ,\nonumber\\
&&\hskip-0.5cm H_5 = 1 + \frac{r_5^2}{r^2} , \qquad H_1 = 1 + \frac{r_1^2}{r^2} ,
\qquad H_P = 1+\frac{r_P^2}{r^2} ,
\labell{five-dim-einstein-metric}
\end{eqnarray}
and the previous case corresponded to $r_1^2=r_5^2=r_P^2=Q$. (We will
shortly identify the origin of the different scales in
equations~\reef{scales}.) Now we can write a more general formula for
the entropy--area relation:
\begin{eqnarray}
S =  \frac{\mathcal{A}}{4 G_5} =  \frac{2 \pi^2}{4 G_5} \sqrt{ r_1^2 r_5^2 r_P^2 -
  \frac{J^2}{4}}\ ,
\labell{generalentropy}
\end{eqnarray}
with an obvious bound on the angular momentum: $ r_1^2 r_5^2 r_P^2 -
  {J^2/4}$. A quick computation shows that this is the same bound
  which, when violated, gives CTCs above the horizon.
  
  Once we begin to work with this more general case, we must note that
  are three scalar fields (we won't give their forms here) which can
  be chosen as corresponding to the dilaton, the radius of a circle
  and the volume of a four--surface $\mathcal{M}$; these are the five
  extra dimensions taking us back to ten dimensions. Now $\mathcal{M}$
  can be $T^4$ or K3, but very soon we will be focusing on the case of
  K3.  We will wrap D5--branes on $\mathcal{M}$ and combine them with
  D1--branes transverse to~$\mathcal{M}$, and subsequently the
  resulting string--like object will be wrapped upon the circle. There
  will be non--trivial momentum in that circle. The gauge field
  separates into two independent one--forms and a two--form potential
  (the latter arrived at by $D{=}5$ Hodge dualisation) and these
  naturally have an interpretation as a Kaluza--Klein gauge field from
  reduction on the circle, the unwrapped part of the R--R 2--form
  gauge field coming from wrapping the D1--branes, and the ($D{=}10$)
  Hodge dual of the R--R 6--form coupling to the D5--branes.
  
  The full ten dimensional geometry is given, in string frame,
  by\cite{Herdeiro}:
\begin{eqnarray} 
ds^2 & = & H_5^{-1/2} H_1^{-1/2} \Big( - dt^2 + \frac{r_P^2}{r^2}(dt-dz)^2 + dz^2
+ \frac{J}{r^2}(\sin^2\theta ~d\phi_1 - \cos^2\theta d\phi_2)( dz - dt) \Big) \nonumber\\
& & + H_5^{-1/2} H_1^{1/2} V^{1/2} ds_{\mathcal{M}}^2 + H_5^{1/2} H_1^{1/2}
( dr^2 + r^2 ~d\Omega_3^2 ) \:,\labell{rotating-D5D1-metric}
\end{eqnarray}
Here, $z$ parameterises our circle, and $ds_{\mathcal{M}}^2$ is the
metric on the manifold $\mathcal{M}$, of unit volume. We denote the
volume element on it as $\varepsilon_{\mathcal{M}}$. $\mathcal{M}$'s
volume varies with the radial coordinate of the transverse space as
\begin{equation}
V(r)= V {H_1 \over H_5}\ ,
\labell{volume}
\end{equation} reaching the
asymptotic value $V$ at spatial infinity. The dilaton and R--R
potentials are\cite{Herdeiro}:
\begin{eqnarray}
e^{2\Phi} & = & g_s^{2} \frac{H_1}{H_5} , \nonumber \\
C^{(6)} &=& g_s^{-1} H_5^{-1} ~dt \wedge dz \wedge
\varepsilon_{\mathcal{M}}
 + \frac{J}{2r^2} H_5^{-1} ( \sin^2\theta
d\phi_1 - \cos^2\theta d\phi_2 ) \wedge dz \wedge 
\varepsilon_{\mathcal{M}}\ , \nonumber \\
C^{(2)} &=& g_s^{-1} H_1^{-1} ~dt \wedge dz + \frac{J}{2r^2} H_1^{-1} (
\sin^2\theta d\phi_1 - \cos^2\theta d\phi_2 ) \wedge dz\ ,
\labell{RR-potential-rotating}
\end{eqnarray}
and we remind the reader that the harmonic functions pertaining to
D5--, D1--branes and the pp--wave (representing the momentum in the
$z$--circle) are given in equations~\reef{five-dim-einstein-metric}.
The scales given in those equations are set by the string coupling
$g_s$, string length $\ell_s$, $\mathcal{M}$'s asymptotic volume $V$,
and the radius, $R_z$, of the circle parameterised by $z$:
\begin{equation} 
r_5^2 = g_s \ell_s^2 Q_5 \ , \quad r_1^2 = g_s \ell_s^2
\frac{V_\star}{V} Q_1 \ , \quad r_P^2 = g_s^2 \ell_s^2
\frac{V_\star}{V} \frac{\ell_s^2}{R_z^2} Q_P \ ,\labell{scales}
\end{equation}
where $V_{\star}=(2\pi\ell_s)^4$.  $Q_1$ and $Q_5$ are integer amounts
of the basic R--R two--form and R--R six--form charges present in the
system.  Note also that $Q_P$ is an integer, parameterising the
discrete amounts of momentum that we can have in the compact direction,
$z$.

Later we will define the ten dimensional Einstein frame metric
$G_{MN}$, in terms of the string frame metric $g_{MN}$ in equation
(\ref{rotating-D5D1-metric}) by $G_{MN} = e^{-\Phi/2}g_{MN}$. Note
also that the ten dimensional Newton constant $G_{10}$ is given by the
relation $16 \pi G_{10} =(2\pi)^7 \ell_s^8 g_s^2$.  The five
dimensional Newton's constant is related to it by $G_5 = G_{10} / 2
\pi R_z V$.

In the case when $\mathcal{M}$ is K3, we must be careful\cite{jm}.
Wrapping a D5--brane on K3 induces precisely minus one units of
D1--brane charge\cite{Bershadsky:1995sp}.  So defining $N_5$ and $N_1$
to be the numbers of D5-- and D1--branes, the charges in
equation~(\ref{scales}) are
\begin{equation} Q_5 = N_5 \ , \qquad Q_1 = N_1 - N_5 \ .
\end{equation}

The configuration preserves 1/8 of the original IIB supersymmetry: 1/2
is broken by having D5--branes, another 1/2 by wrapping them on K3
({\it or} combining them with D1--branes), and finally a pp--wave in
the $z$--direction with purely right--moving momentum excited breaks
1/2 of the remaining supersymmetry.  Rotation does not break an extra
amount of supersymmetry, but for the solution to be regular the linear
combination of angular momenta in $\phi_1$ and $\phi_2$ directions
should vanish\cite{Breckenridge:1996is, Tseytlin:1996as}, and
$J_{\phi_1} = - J_{\phi_2}$.

\section{Two Dimensions}
\label{cft}

There is a $(1+1)$--dimensional superconformal field theory living on
the world--volume of the string--like intersection of the D5--branes
and D1--branes with a number of interesting properties relevant to the
spacetime physics. We present it here, following closely the original
reference\cite{Breckenridge:1996is}. The theory lives on the cylinder
$S^1_z\times \IR$, and has four supercharges. It has states coming
from the massless strings stretching between the various D1-- and
D5--branes. The usual 1--5 and 5--1 strings give the net contribution
to the degrees of freedom ($N_B=N_F=4Q_1Q_5$), giving a central charge
$c=N_B+N_F/2=6Q_1Q_5$. The R--symmetry of this theory is the $SO(4)$
isometry of the $\IR^4$ transverse to the intersection.  The relation
between the coordinates $(x_1,x_2,x_3,x_4)$ of this $\IR^4$ and the
coordinates $(r,\theta,\phi_1,\phi_2)$ we have been using so far is:
\begin{eqnarray}
&&x_1=r\sin\theta\cos\phi_1 \ ,\qquad x_3=r\cos\theta\cos\phi_2 \ ,\nonumber\\
&&x_2=r\sin\theta\sin\phi_1 \ ,\qquad x_4=r\cos\theta\sin\phi_2 \ .
\labell{coordchange}
\end{eqnarray}
The maximal Abelian subgroup of this $SO(4)$ is $U(1)_{12}\times
U(1)_{34}$ corresponding, say, to rotations in the two orthogonal
planes indicated by the labelling.  These also give the two
independent angular momenta $(J_{12}, J_{34})$ that a point--like object
in five dimensions are allowed to have. The BPS condition on the
spacetime solution allows only one linear combination of these two
angular momenta to be non--zero, and there is an analogous situation
in the two dimensional CFT corresponding to a condition on the allowed
masses and charges of states ($J_{12}, J_{34}$) which are excited
there, restricting them to be BPS states.

The convention is that the linear combination $J_{12}-J_{34}$ is
called $J_L$ and the other is $J_R=J_{12}+J_{34}$. We have $J_L=0$ and
the non--zero $J_R$ is related to the spacetime $J$ by
\begin{equation}
J_R=\frac{\pi}{4G_5}J\ .
\labell{convert}
\end{equation}
In the conformal field theory on the cylinder, the $L_0$ energy
eigenvalue of a state is $Q_P$. Unitarity, and an examination of the
superconformal algebra requires that the energy (conformal weight) of
a state of R--charge $J_R$ is bounded: $Q_P\geq 3J_R^2/(2c)$, which
translates into our bound from before arising from the spacetime
entropy~\reef{generalentropy}, or absence of spacetime CTCs:
\begin{equation}
Q_1Q_5Q_P\geq \frac{J_R^2}{4}\quad \Longleftrightarrow \quad r_1^2r_5^2r_P^2\geq \frac{J^2}{4}\ ,
\end{equation}
where we have converted the charges using the relation in
equations~\reef{scales} and~\reef{convert}.

The entropy of the black hole for large charges $Q_1,Q_5$ is just
given by the logarithm of the standard formula (see {\it e.g.},
ref.\cite{gsw}) for the asymptotic level density of states, $d(n,c)$,
of charge $J_R$ as the level $n=Q_P-3J_R^2/(2c)$ becomes large:
\begin{equation}
d(n,c)\simeq \exp\left(2\pi\sqrt{\frac{nc}{6}}\right) \qquad \Longrightarrow
\qquad S=2\pi\sqrt{Q_1Q_5Q_P-\frac{J_R^2}{4}}\ ,
\end{equation}
which easily converts (using equations~\reef{scales}
and~\reef{convert}) to the entropy~\reef{generalentropy} computed from
the supergravity.

\section{The Basic Enhan\c con Locus}
\label{section-rotating-enhancon}

The ten dimensional solution exhibits a naked repulson
singularity\cite{repulsive}, at the place where the K3 volume shrinks
to zero\cite{jpp}. This unphysical behaviour is repaired by the
enhan\c con mechanism~\cite{jpp}: this repulson part of the geometry
must be a supergravity artifact, since new degrees of freedom must
have come to play at the radius where the K3 volume $V(r)$ reached the
special value $V_{\star} = (2 \pi \ell_s)^4$. A quick computation
shows that this ``enhan\c con'' radius is precisely the same as in the
non--rotating black hole case of ref.~\cite{jm}:
\begin{equation} 
r_{\rm e}^2 = g_s \ell_s^2 \frac{V_\star}{(V - V_{\star)}} (2N_5 - N_1)
,\labell{r_e-rotating}
\labell{eradius}
\end{equation}
where negative $r_{\rm e}^2$ means that the enhan\c con lies inside
the event horizon (located at $r^2=0$). This is the case of
\mbox{$2N_5<N_1$}. 

Note that the enhan\c con locus is spherically symmetric. This is a
feature of the geometry which is very useful in much of our analysis,
allow for much simplification, as we will comment further later. The
harmonic functions $H_{1,5}$ in equation~\reef{volume} have no angular
dependence at all, despite the rotation.

The above radius~\reef{eradius} is the enhan\c con radius uncovered in
ref.\cite{jpp}, where the tension of a wrapped D5--brane would fall to
zero. Such a brane cannot proceed further inside the geometry as its
tension would go negative. Its motion ends at $r_{\rm e}^2$, and such
probes can form a shell of tensionless branes at this radius. We'll
write a new supergravity solution for this possibility later.  As
observed in ref.\cite{jm}, the presence of other species of brane,
allowing for the formation of D5--D1 bound states, gives a much richer
behaviour. There are other enhan\c con loci corresponding to the place
where these bound states can become massless and can proceed no further
into the interior. Let us study these probes in the background of this
geometry next.

\section{Probing the geometry}
\label{probing-section}

Let us probe the geometry with a bound state of $n_5$ D5--branes and
$n_1$ D1--branes. The effective world--sheet  action for such a composite
brane probe is\cite{jm}:
\begin{eqnarray}
S & = & - \int_{\Sigma} d^2\xi e^{-\Phi} ( n_5 \tau_5 V(r)
+ (n_1-n_5)\tau_1)(-\det g_{ab})^{1/2} \nonumber \\
&& \quad + n_5 \mu_5 \int_{\Sigma \times K3} C^{(6)} +
(n_1-n_5)\mu_1 \int_{{\mathcal M}_2}C^{(2)} \ ,
\end{eqnarray}
where $n_5$ and $n_1$ denote the number of D5-- and D1--branes we have
assembled to make up the probe.  We have called the world--sheet
$\Sigma$. It is assumed that $n_5 \ll N_5$ and $n_1 \ll N_1$.

In the static situations the terms from the Wess--Zumino part of the
effective action (second line) are cancelled by contributions from the
Dirac--Born--Infeld part (first line), so that only the kinetic term
remains. This happens due to the BPS condition --- the ``electric''
Coulomb repulsion is balanced by the gravitational attraction. In the
rotating case one would expect that the analogue of frame dragging
effects give some additional terms to the DBI part of the effective
action. Happily, the R--R potential in
equation~(\ref{RR-potential-rotating}) is endowed with additional
``magnetic'' components, and so the WZ part of the action also gets an
extra contribution. They cancel, as we see immediately below, and this
is a consequence of the fact that we have a BPS system.

We adopt a static gauge, defining the coordinates on the probe brane
world--volume $\Sigma$ to be $\xi^0=t, \xi^1=z$. The probe brane can
move in the transverse directions $x^i=x^i(t)$, $x^i = (r, \theta,
\phi_1, \phi_2)$, but is frozen on K3.  Pulling back the string frame
metric to the probe world--volume and expanding the square root gives
\begin{eqnarray} 
\mathcal{L}_{\rm DBI} & = & - 
( n_5 \tau_5
H_5^{-1}V + (n_1-n_5)\tau_1 H_1^{-1}) \Big(1 + \frac{J}{2r^2}
\big(\sin^2\theta \dot{\phi}_1 - \cos^2\theta
\dot{\phi}_2 \big) \Big) \nonumber \\
&& \qquad \qquad \quad - \frac{1}{2}( n_5 \tau_5 H_1 V +
(n_1-n_5)\tau_1 H_5) H_P v^2\ 
,\labell{S_DBI}
\end{eqnarray}
where we have assumed slow motion of the probe, \textit{i.e.}
\begin{equation}
v^2 = \dot{r}^2 + r^2 \left(\dot{\theta}^2 + \sin^2\theta
\dot{\phi_1}^2 + \cos^2\theta \dot{\phi_2}^2 \right)
\labell{vee}
\end{equation}
is taken be small. Similarly the WZ part of the effective action is
given by
\begin{equation} 
\mathcal{L}_{WZ} = 
(n_5 \tau_5 f_5^{-1} V +
(n_1-n_5)\tau_1 f_1^{-1} ) \Big( 1 + \frac{J}{2r^2} \big(
\sin^2\theta \dot{\phi}_1 - \cos^2\theta \dot{\phi}_2 \big) \Big)\
.\labell{S_WZ}
\end{equation}
As we can see, not only do the potential terms of the DBI and WZ parts
of the effective action cancel, but  the $J$ terms linear in
angular velocity  cancel as well ---the effects of gravitational
frame dragging are neatly balanced by the ``magnetic'' force induced
by rotation.\footnote{Otherwise it has been noticed in {\it e.g.} ref.
  \cite{Cai:1999ad}, that in the non-extremal rotating D3--brane
  background the effective action of a probe D3--brane contained a
  term proportional to the probe angular velocity.}

Putting equations~(\ref{S_DBI}) and (\ref{S_WZ})
together we find that only the kinetic term survives in the
effective Lagrangian
\begin{equation} 
{\cal L} = \frac{1}{2}( n_5 \tau_5 H_1 V + (n_1-n_5)\tau_1 H_5) H_P
v^2.\labell{L_static}
\end{equation}
The prefactor of the kinetic energy gives the effective tension of
the probe. This tension is positive as long as
\begin{equation}
r^2 > r_{\rm e}^2[n_1,n_5]=g_s \ell_s^2 V_\star \frac{(2N_5 - N_1) n_5 - N_5 n_1}{(V-
V_\star)n_5 + V_\star n_1} \ , \labell{probe-r_e-rotating}
\end{equation}
while the locus of the vanishing tension indicates the position of the
enhan\c con for the $(n_1,n_5)$ bound state, generalising the
expression given in equation~\reef{eradius}, which is the case
$n_1=0$.  Note that at these radii, the volume of K3 is below
$V_{\star}$.

We will find out in the next section \ref{section:rotating-excision}
that the lower bound where the tension vanishes agrees perfectly with
the results of supergravity computations, where we build new
geometries containing the shell formed by bringing up probes to the
locus of points where their tension vanishes.

For a probe made up of D1--branes only, ({\it i.e.}, $n_5=0$), the
tension remains positive everywhere and hence D1--brane probes can
make their way freely down to $r=0$ without being forced to stop by an
enhan\c con locus. For the case $n_5=n_1$, the tension of such a probe
is also positive everywhere.  This is how we can imagine constructing
a black hole of arbitrary charges. The result of the previous
subsection might have suggested that we cannot successfully bring
individual D5--branes up to the horizon at $r^2=0$, depending upon the
charges already present. This result for the $n_1=n_5$ case gives us
an avenue around this\cite{jm}, as we can move the D1--branes already
present to the D5--brane enhan\c con locus, bind them with the
D5--branes already there, and then move them in as bound states,
thereby constructing a hole with arbitrary charges of one's choice.

\section{Excision}
\label{section:rotating-excision}

Although by the processes described immediately above we can take
D5--branes inside the enhan\c con radius to construct the geometry
described in solution~\reef{rotating-D5D1-metric}, we learn that there
can be geometries where branes with a certain set of charges
$(n_1,n_5)$ can remain ``hung up'' at a specific radius. They form a
shell at that radius, made of zero tension branes. In fact, using
purely supergravity techniques, we can check that this is a consistent
picture by building the suggested geometry. We glue two geometries
together at some radius $r=r_{\rm i}$, with excess charge in the outer
geometry and check what the conditions at the junction between the two
suggest for the physics there. We confirm that it is consistent with
the probe picture, with particular attention given to the case $r^2_{\rm
  i}=r^2_{\rm e}[n_1,n_5]$. The procedure goes through pretty much along
the same lines as in the static case studied in refs.\cite{jmpr,jm}.
The shells we have here are all spherically symmetric, despite the
rotation. As already stated above, this is because the loci of equal
K3 volumes (given by equation~\reef{volume}) are spherical, a very
special feature of the rotating geometry. This is in contrast to other
recent cases studied in the literature, where the BPS enhan\c con
shells are highly non--spherical\cite{oblate,Astefanesei:2001ki}, and
even disconnected\cite{laurthesis}.

We keep the total number of D--branes as measured at infinity as $N_1$
and $N_5$, as before. Now, however, certain numbers, $\delta N_5$ and
$\delta N_1$, of D5--branes and of D1--branes respectively are located
on the shell at $r_{\rm i}$, while $N'_5 = N_5 - \delta N_5$ of
D5--branes and $N'_1 = N_1 - \delta N_1$ of D1--branes are located in
the interior.  The solution describing the interior region has the
same form as equation~(\ref{rotating-D5D1-metric}) above, only the
harmonic functions $H_1$, $H_5$ are now substituted by
\begin{equation} h_1 = 1 + \frac{r_1^2 -
 \tilde{r}_1^2}{r_{\rm i}^2} + \frac{\tilde{r}_1^2}{r^2},
 \qquad
h_5 = 1 + \frac{r_5^2 - \tilde{r}_5^2}{r_{\rm i}^2}
 + \frac{\tilde{r}_5^2}{r^2} ,
\labell{h_1-h_5}
\end{equation} with the scales
\begin{equation} \tilde{r}_1^2 = g_s \ell_s^2 \frac{V_\star}{V} Q'_1 \ ,
\qquad \tilde{r}_5^2 = g_s \ell_s^2 Q'_5 ,
\labell{tilde-r_1-rotating}
\end{equation}  being
proportional to the number of branes inside, \textit{i.e.} $Q'_5 =
N'_5$, $Q'_1 = N'_1 - N'_5$. We keep $Q'_1 \geq 0$ to avoid a repulson
singularity. The functions $h_1$ and $h_5$ in
definition~\reef{h_1-h_5} are chosen so that the metric of the
corrected solution is continuous at $r_{\rm i}$. The discontinuity in
the extrinsic curvature on the junction surface at $r_{\rm i}$ has an
interpretation as the surface stress-energy tensor of this thin
shell\cite{MTW:book,lanczos}.  After some algebra, we find
that the stress-energy tensor of the gluing surface is of the same
form as in the static case\cite{jm}, but now with additional
$(t,\phi_i)$ and $(z,\phi_i)$ terms:
\begin{eqnarray}
S_{\mu\nu} & = & \frac{1}{2 \kappa^2 \sqrt{G_{rr}}} \left( \frac{H_1^\prime}{H_1} + \frac{H_5^\prime}{H_5} - \frac{h_1^\prime}{h_1} -  \frac{h_5^\prime}{h_5}\right) G_{\mu\nu} ,
\nonumber \\
\labell{junction-stress-t-phi}
S_{\mu\phi_i} & = & \frac{1}{2 \kappa^2 \sqrt{G_{rr}}} \left( \frac{H_1^\prime}{H_1} + \frac{H_5^\prime}{H_5} - \frac{h_1^\prime}{h_1} -  \frac{h_5^\prime}{h_5}\right) G_{\mu\phi_i} ,
\nonumber \\
S_{ij} & = & 0 , \nonumber \\
S_{ab} & = & \frac{1}{2 \kappa^2 \sqrt{G_{rr}}} \left(\frac{H_5^\prime}{H_5} -  \frac{h_5^\prime}{h_5}\right) G_{ab} ,
\end{eqnarray}
where indices $\mu$, $\nu$ denote the $t$ and $z$ directions, $a$, $b$
denote the K3 directions, $i$, $j$ denote the angular directions along
the junction $S^3$. The Einstein frame metric $G_{MN}$, natural in
this computation, is related to the string frame metric $g_{MN}$ in
equation (\ref{rotating-D5D1-metric}) by $G_{MN} = e^{-\Phi/2}g_{MN}$,
and $2 \kappa^2 = 16 \pi G_{10}$.

The tension along the angular directions vanishes, since despite
rotation we still have a BPS system that does not need some force
between the branes to support the shell at arbitrary radius. In the
K3 directions the tension depends only on the harmonic functions of
the D5--branes as only they wrap these directions.  Finally in the $t$
and $z$ directions as well as $t-\phi_i$ and $z-\phi_i$ directions the
surface stress-energy is proportional to a tension
\begin{equation} 
T_{\rm eff} = \frac{1}{2 \kappa^2 \sqrt{G_{rr}}} \left(
\frac{H_1^\prime}{H_1} + \frac{H_5^\prime}{H_5} -
\frac{h_1^\prime}{h_1} -  \frac{h_5^\prime}{h_5}\right)\labell{surface-tension}
\end{equation}
These results are consistent with what one would expect from the fact
that the shell is built of D5-- and D1--brane sources \cite{jmpr}.

If there are no D1--branes on the shell ($\delta N_1 = 0$), the
tension (\ref{surface-tension}) vanishes precisely at the basic
enhan\c con radius given in equation~(\ref{r_e-rotating}).
Alternatively if some D1--branes stay on the shell, the tension is
positive down to
\begin{equation} 
\tilde{r}_{\rm e}^2 = g_s \ell_s^2 V_\star \frac{(2N_5 - N_1) \delta N_5
- N_5 \delta N_1} {(V- V_{\star}) \delta N_5 + V_{\star} \delta
N_1},\labell{tildere}
\end{equation}
and we note that $\tilde{r}_{\rm e}^2 < r_{\rm e}^2$.  Satisfyingly,
this lower bound where the tension vanishes agrees perfectly with the
$(n_1,n_5)$ probe computation result for the enhan\c con radii $r_{\rm
  e}^2[n_1,n_5]$ given in equation~(\ref{probe-r_e-rotating}) if one
substitutes $\delta N_1 \rightarrow n_1$, $\delta N_5 \rightarrow
n_5$. We see that the consistency conditions derived here and the
probe results of the previous section are in perfect agreement with
each other.

We note that as we replace the geometry inside of the enhan\c con
radius with a repaired geometry given by the harmonic functions
(\ref{h_1-h_5}) the running K3 volume in the interior is now
\begin{equation} V(r) = \frac{h_1}{h_5} V_{\star} \ ,
\labell{Vinside}
\end{equation} and in particular at the horizon
\begin{equation} V(r=0) = \frac{\tilde{r}_1^2}{\tilde{r}_5^2} V_{\star} = \frac{N_1' - N_5'}{N_5'} V_{\star}\ .
\labell{Vhorizon}
\end{equation} The volume at the enhan\c con radius is still $V_{\star}$, but
now we have a possibility that inside, {\it i.e.} for
$r<\tilde{r}_{\rm e}^2$, it can actually grow larger.  Therefore means
that some of the D5--branes can actually pass the enhan\c con radius
and move in as their tension does not become negative in the process.
This is subject to the condition $N'_1 > 2 N'_5$, apparent also from
equation~(\ref{Vhorizon}), which shows $V(r=0) > V_{\star}$ if the
condition holds.  In fact, in the limit $N'_1 = 2 N'_5$ the K3 volume
is $V(r) = V_{\star}$ uniformly everywhere in interior up to the
enhan\c con radius.

Notice that at arbitrary excision radius, the $S_{\mu\phi_i}$
components of the surface stress--energy tensor depends on
$G_{\mu\phi_i}$ and hence on $J$, while at the enhan\c con radius
these components of stress--energy vanish. This indicates that the
zero--tension enhan\c con shells also have vanishing angular momentum.
(We shall see that this is consistent with a probe's motion in the
geometry in the next section: If they have non--zero angular momentum
the probe computations show that they cannot stay in the shell).  This
result is reminiscent of one of the already mentioned special features
that these black hole solutions have\cite{Gauntlett:1998fz} which is a
vanishing angular velocity of the horizon. This analogy should not be
stretched too far, however, since in that case, the vanishing is taken
to be a result of a cancellation of opposite dragging effects: there
is a opposite sense of rotation between the two sides of the horizon.
Here, the senses of rotation on either side of a generic enhan\c con
shell are the same.

\section{The Second Law and CTCs}
\label{entropy-section}

So, as we have already computed, the entropy of our five dimensional
black hole given in equation~\reef{five-dim-einstein-metric} is given by:
\begin{eqnarray}
S  =  \frac{\mathcal{A}}{4 G_N^{(5)}} 
   =  \frac{2 \pi^2}{4 G_N^{(5)}} \sqrt{ r_1^2 r_5^2 r_P^2 -
   \frac{J^2}{4}}
 =  2 \pi \sqrt{(N_1 - N_5) N_5 Q_P - \frac{J^2_R}{4 }}
\labell{rotating-black-hole-entropy}
\end{eqnarray}
where in the last term we have written it in terms of the actual {\it
  number} of each type of brane, as opposed to the net charges. The
essential novelty here is the presence of minus signs in the part
involving the charges, which is due to wrapping our branes on K3
instead of $T^4$, where we would have had simply $N_1 N_5 Q_P$.

The key point, observed first in ref.\cite{jm}, is the fact that for a
probe with the correct choice of D1-- and D5--brane charges, bringing
it to the horizon would {\it reduce} the entropy.  This process can be
done as slowly as we like (given our probe computations in
section~\ref{probing-section}), and the resulting adiabaticity gives
us a very clear violation of the Second Law of Thermodynamics. The
area theorem assures us that this cannot happen, and the novelty we
wish to observe is that the enhan\c con mechanism operates precisely
to ensure that the area theorem ---and hence the Second Law--- is
inviolate.

The first order change in the entropy is:
\begin{eqnarray}
\delta S&=&2\pi^2 S^{-1}\left\{Q_P\left(N_5\delta N_1 +(N_1-2N_5)\delta
  N_5\right)-\frac{J_R\,\delta J_R}{2}\right\}\nonumber \\
&=&
 2\pi^2 S^{-1}\left\{Q_P\left(n_1 N_5+n_5(N_1-2N_5)\right)-\frac{J_R\,\delta J_R}{2}\right\}\ ,
\labell{change}
\end{eqnarray}
and we have inserted $\delta N_1=n_1$ and $\delta N_5=n_5$ for the
charges on the probe. We have neglected the change in $Q_P$ since it
could only decrease $S$ by itself decreasing. This obviously cannot be
achieved with a probe while retaining the saturation of the BPS
condition which takes us out of the class of processes which we wish
to consider.

Let us consider first probes with no angular momentum, and so we set
$\delta J_R=0$. In such cases then, the key observation is that the
entropy change is negative if $n_1 N_5+n_5(N_1-2N_5)$ is
negative, which is the same condition for an enhan\c con locus to
appear above the horizon (see equation~\reef{tildere}
or~\reef{probe-r_e-rotating}), stopping that particular 
probe from reaching the horizon. This is how the enhan\c con mechanism
protects the Second Law from dangerous probes.

We can of course have probes with non--zero angular
momentum\footnote{We are of course assuming that the charges are such
  that $4(n_1-n_5)n_5q_P\geq j_R^2$ on an individual probe, where
  $j_R$ is its angular momentum and $q_P$ is its $z$--momentum,
  otherwise the theory on the probe is non--unitary at the outset. See
  section~\ref{cft}.}, $j_R$.  In fact, we must, or we cannot
actually construct the BPS black hole with non--zero angular momentum
at all. There are again two cases: Dangerous probes, for which the
R--R charges are such that they can reduce $(N_1-N_5)N_5Q_P$, and
probes for which the R--R charges cannot decrease $(N_1-N_5)N_5Q_P$.
For the latter sort, there is no novel enhan\c con physics to be
found, since the physics there is exactly the same as in the case of
having made a hole by wrapping on $T^4$. The area theorem is protected
in the usual way and we consider them no further.

For the dangerous sort of angular momentum carrying probes, we must
consider the ways in which they can bring angular momentum into the
hole: {\it (a)} They can have intrinsic angular momentum $j_R$, which
means that their world--sheet CFT of section \ref{cft} has states with
non--trivial R--charge, {\it (b)} they can have non--zero impact
parameter contained in the geometry of their approach, giving them
``orbital'' angular momentum, and {\it (c)} they can have some mixture
of the two previous cases.  The effective Lagrangian for a probe's
slow motion in the geometry is given by equation~\reef{L_static},
where $v^2$ is given by equation~\reef{vee}.  The orbital angular
momenta  are the conjugate momenta to $\phi_1$ and $\phi_2$:
\begin{eqnarray}
j_{\phi_1}\equiv\frac{\partial {\mathcal{L}}}{\partial
  {\dot\phi_1}}=r^2F(r)\sin^2\theta {\dot \phi_1}\ ,\qquad
j_{\phi_2}\equiv\frac{\partial {\mathcal{L}}}{\partial
  {\dot\phi_2}}=r^2F(r)\cos^2\theta{\dot\phi_2}\ ,
\labell{momenta}
\end{eqnarray}
 where
\begin{equation}
F(r)=( n_5 \tau_5 H_1 V + (n_1-n_5)\tau_1 H_5) H_P\ .
\labell{eff}
\end{equation}
As these momenta are conserved, we can initially reduce our problem to
a two dimensional one in $r$ and $\theta$, with an effective potential
set by the angular momenta:
\begin{eqnarray}
\mathcal{L}&=&\frac{1}{2}F(r)({\dot
  r}^2+r^2{\dot\theta}^2)+\frac{1}{2r^2F(r)}\left(\frac{j_{\phi_1}^2}{\sin^2\theta}+\frac{j_{\phi_2}^2}{\cos^2\theta}\right)
  \ . \labell{effective}
\end{eqnarray}
The case {\it (a)} above corresponds to vanishing $j_{\phi_1}$ and
$j_{\phi_2}$, for which it is clear again that we have the same
discussion as above: Either $F(r)=0$ above the horizon and the enhan\c
con mechanism forbids further approach, or the mechanism is not
relevant and the area is understood to be non--decreasing in the
conventional fashion.

Cases {\it (b)} or {\it (c)} are no longer BPS. We need to
have equal angular momenta in the 1--2 and 3--4 planes and to satisfy
the BPS condition, with the angular momentum tensor $M^{\mu\nu}$ in the form:
\begin{equation}
M\sim\pmatrix{\phantom{-}0&j&\phantom{-}0&0\cr -j&0&\phantom{-}0&0\cr\phantom{-}0&0&\phantom{-}0&j\cr\phantom{-}0&0& -j&0}\ . 
\labell{required}
\end{equation}
We can place such an equality condition on the orbital angular momenta
of the probe in the 1--2 and 3--4 planes, but we cannot avoid a
centripetal potential being generated.  There is no assignment of
orbital angular momenta to a single probe which will give a vanishing
potential which is a consequence of the fact that for a single probe
we can easily find a rotation ({\it e.g.}, combining one in the 1--3
with one in the 2--4 plane) to find a frame in which the associated
angular momentum tensor is just\footnote{We thank Rob Myers for
  explicitly pointing this out.}:
\begin{equation}
M\sim\pmatrix{\phantom{-}0&j&0&0\cr -j&0&0&0\cr\phantom{-}0&0&0&0\cr\phantom{-}0&0&0&0}\ .  
\end{equation}
In retrospect, this is clearly a result of the fact that in any
dimension, a particle with conserved orbital angular momentum will
remain constrained to move in a single plane.  In order to get a BPS
configuration with the correct combination of orbital momentum, one
would have to consider two probes, moving such that they have angular
momentum in the two independent planes. Counting parameters and
available rotations shows that it should be possible to achieve an
angular momentum tensor of the form given in equation~\reef{required}.
On the other hand, from this point of view and also from that of the
required probe action, this would seem to be requiring them to be
coupled in an interesting (and apparently non--local) manner. This is
a probe problem which we will not pursue here, since it will take us
beyond the matters we wish to discuss in the present work. In any
event, such a BPS case would then become rather similar to that of
case {\it (a)}, where we have only intrinsic angular momentum, in the
sense that it the probes have the ``wrong'' charges, they will be
subject to the appearance of an enhan\c con locus above the horizon to
stop their approach.

Further to the discussion in the previous paragraph, note that from
equations~\reef{momenta} we see that for fixed angular momenta, we
must ensure that $r^2F(r)\sin^2\theta$ and $r^2F(r)\cos^2\theta$ stay
away from zero, otherwise the velocities ${\dot\phi}_{1,2}$ diverge,
taking us out of the slow--probe limit of
section~\ref{probing-section}. We must avoid the neighbourhood of the
planes ($\theta=0,\pi/2$) lest we violate this, or the probe will
encounter additional forces from the background which we previously
neglected.
  
  Looking further at the non--trivial effective potential given in
  equation~\reef{effective}, (and even staying sufficiently far away
  from the special planes $\theta=0,\pi/2$) we see that there is an
  infinitely repulsive wall at the enhan\c con locus for such cases of
  non--zero impact parameter, naturally induced by the vanishing
  kinetic term there. Near there, the velocities cannot be small, and
  so there will be further terms which we neglected which introduce
  more forces on the probe due to the background. Unless there is a
  remarkable conspiracy, it is unlikely that these terms can soften
  this infinite repulsion at the enhan\c con locus, and so our result
  is consistent with the conclusions reached for a BPS approach.

Finally, we note that the impossibility of adding certain charges on
D--brane probes to the hole in order to reduce the entropy also means
something for the occurrence of CTCs: If we start with a hole with no
CTCs above the horizon, we simply cannot introduce a probe to the
black hole which will create a CTC, since this would require $S$ in
equation~\reef{rotating-black-hole-entropy} to reduce, and the enhan\c
con mechanism forbids that. This is in accord with existing discussion
in the literature\cite{Gibbons:1999uv} about not being able to
manufacture CTCs (at least for this class of geometries) by a physical
process.  As we pointed out in section~1, this does not rule out
considering a CTC--endowed object with this geometry, since they do
not owe their existence to the presence of K3, while the enhan\c con
(in this example) does.

\bigskip
\bigskip
\bigskip

{\bf Acknowledgements:} Some of the computations in sections~5, 6 and
7 were carried out simultaneously with Lisa M. Dyson. She declined to
be listed as a co--author on this paper, despite appearing as such on
the first released version. We do not know why. Nevertheless we fully
recognise ---and thank her for--- her computational contribution. We thank
Carlos Herdeiro for helpful comments during the initial stages of this
research, and Robert C.  Myers for crucial comments on an earlier
version of this manuscript.  We thank Douglas Smith and Paul Townsend
for comments. LJ is grateful to the CPT, University of Durham, where
he worked on this project as a PhD student, while partially supported
by a British ORS Scholarship, by a Durham University Studentship and
also by the Estonian Science Foundation Grant No 5026.  Some of CVJ's
research was supported by an EPSRC grant.  This paper is report
numbers FSU TPI 06/02 and DCPT-02/77.

\end{document}